\documentclass{article}
\usepackage{epsfig}
\usepackage[numbers,sort&compress]{natbib}
\usepackage{rotating}
\usepackage{graphicx}

\usepackage{graphicx}
\usepackage[english]{babel}
\usepackage{listings} 
\newcommand{\bfr}{\begin{flushright}}
\newcommand{\efr}{\end{flushright}}

\begin{document}
\title{\LARGE \bf Inclusive charged light di-hadron
production at 7 and 13 TeV LHC \\ in the full NLA BFKL approach
\thanks{Presented at the Low x workshop, June 13-18 2017, Bari, Italy}%
}
\author{\vspace{-0.20cm} \\
F.G. Celiberto$^{1,2}$\footnote{Speaker; francescogiovanni.celiberto@fis.unical.it}\hspace{0.1cm}, 
D.Yu. Ivanov$^{3,4}$, 
B. Murdaca$^2$ and 
A. Papa$^{1,2}$
\\~\\
{\small
\centerline{${}^1$ {\sl Dipartimento di Fisica, Universit\`a della Calabria, Arcavacata di Rende, I-87036 Cosenza, Italy}}
}
\\
{\small
\centerline{${}^2$ {\sl Istituto Nazionale di Fisica Nucleare, Gruppo collegato di
Cosenza, Arcavacata di Rende, I-87036 Cosenza, Italy}}
}
\\
{\small
\centerline{${}^3$ {\sl Sobolev Institute of Mathematics, RU-630090 Novosibirsk, Russia}}
}
\\
{\small
\centerline{${}^4$ {\sl Novosibirsk State University, RU-630090 Novosibirsk, Russia}}
}
\smallskip\\
}
\date{\today
}
\maketitle
\begin{abstract}
We give the first phenomenological predictions of cross sections and azimuthal correlations for the inclusive di-hadron production in the full NLA BFKL approach. This process shares the same theoretical framework with the well konwn Mueller--Navelet jet production and can be considered a novel and complementary channel to access the BFKL dynamics at proton colliders.
\\
~
\\
PACS number(s): 
12.38.Bx, 12.38.-t, 12.38.Cy, 11.10.Gh
\end{abstract}

\section{Introduction}

The great amount of data being produced at the Large Hadron Collider (LHC) offers us a unique occasion to study the dynamics of strong interactions in the high-energy limit.
In this kinematical regime, 
the Balitsky-Fadin-Kuraev-Lipatov 
(BFKL) approach~\cite{Fadin:1975cbKuraev:1976geKuraev:1977fsBalitsky:1978ic} 
is the most adequate tool 
to perform the resummation to all orders of the leading (LLA)
and the next-to-leading terms (NLA) of the QCD perturbative series 
which are heightened by powers of large energy logarithms.
\begin{figure}[t]
 \label{fig:di-hadron}
 \centering
 \includegraphics[scale=0.50]{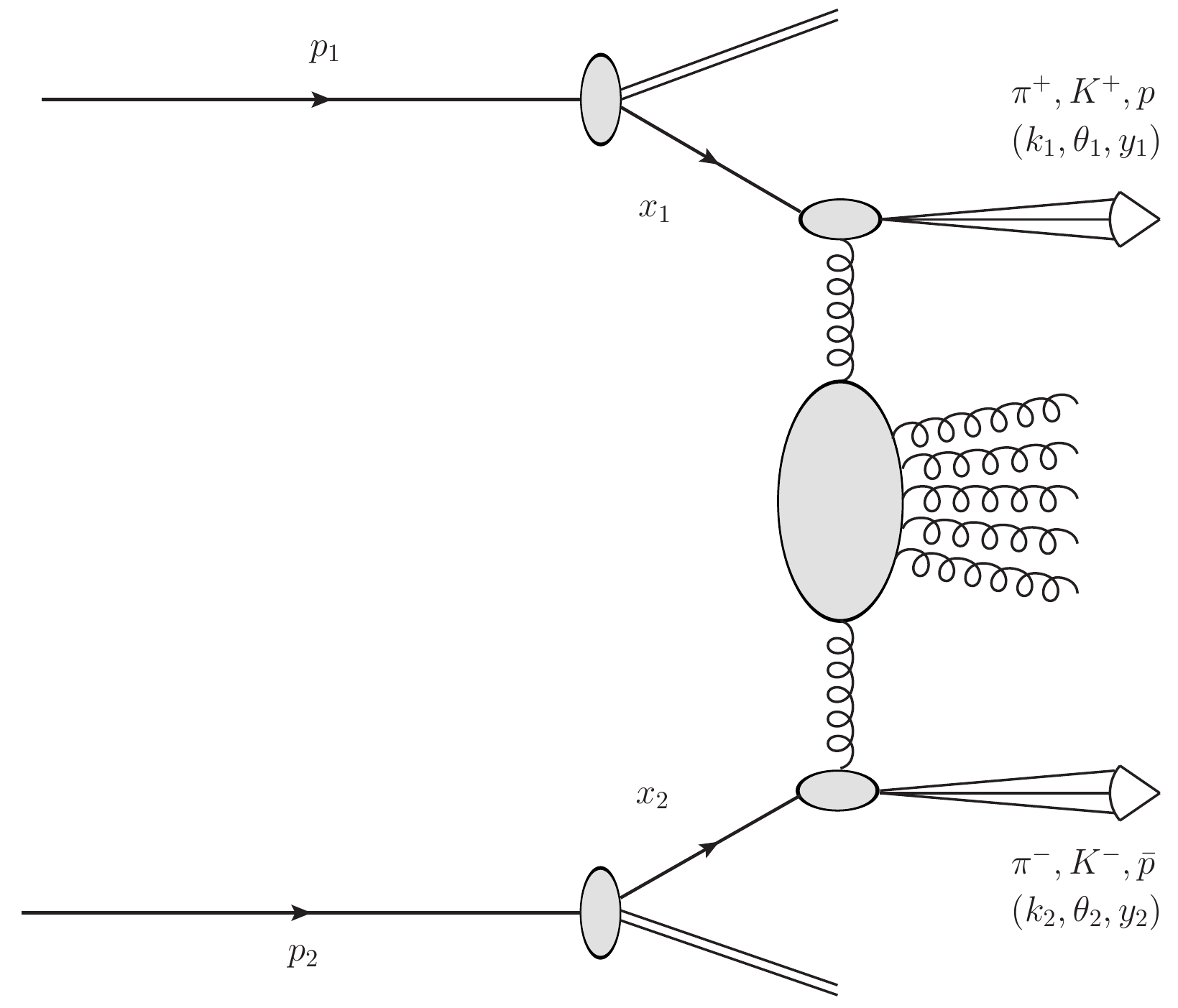}
 \caption{Inclusive di-hadron production process in multi-Regge kinematics.}
\end{figure}
The inclusive hadroproduction of two jets featuring high transverse momenta 
and separated by a large rapidity interval, 
well known as Mueller--Navelet process~\cite{Mueller:1986ey}, 
has been one of the most studied reactions in the last years.
For this process, the BFKL resummation with NLA accuracy  
leans on the compound of two ingredients: the NLA Green's function 
of the BFKL equation~\cite{Fadin:1998py,Ciafaloni:1998gs} and the NLA jet 
vertices~\cite{Bartels:2001ge,Bartels:2002yj,Caporale:2011cc,
               Ivanov:2012ms,Colferai:2015zfa}.
In~\cite{Colferai:2010wu,Angioni:2011wj,Caporale:2012ih,
         Ducloue:2013hia,Ducloue:2013bva,Caporale:2013uva,
         Ducloue:2014koa,Caporale:2014gpa,Ducloue:2015jba,
         Celiberto:2015yba,Celiberto:2016ygs,Chachamis:2015crx}  
NLA BFKL predictions of cross sections and azimuthal angle correlations
for the Mueller--Navelet jet process, observables proposed
in~\cite{DelDuca:1993mn,Stirling:1994he,Vera:2006un,Vera:2007kn}, 
were given, showing a very nice agreement 
with LHC data~\cite{Khachatryan:2016udy}.
In order to further and deeply probe the dynamics behind partonic interactions in the Regge limit, $s\gg |t|$, some other observables, sensitive to the BFKL dynamics, should be considered in the context of the LHC physics program. 
A stimulating option, the detection of three and four jets, well separated in rapidity from each other, was recently suggested in~\cite{Caporale:2015vya,Caporale:2015int} and investigated  in~\cite{Caporale:2016soq,Caporale:2016zkc,Caporale:2016xku,Celiberto:2016vhn}.
In this work a novel possibility, {\it i.e.} 
the inclusive production of two charged light hadrons:
$\pi^{\pm}$, $K^{\pm}$, $p$, $\bar p$ 
featuring high transverse momenta and separated by a large rapidity interval,
together with an undetected gluon radiation emission 
is investigated~(see Fig.~\ref{fig:di-hadron}). 
Likewise for Mueller--Navelet jets, BFKL studies in the NLA accuracy are feasible for this reaction, since NLA expression for the vertex describing the production of an identified hadron was calculated  in~\cite{hadrons}. 
On one side, hadrons can be tagged at the LHC 
at much smaller values of the transverse
momentum than jets, letting us to explore a kinematic range supplementary to the one studied with Mueller--Navelet jets. 
On the other side, this process allow us to constrain not only the parton densities (PDFs) 
for the initial proton, but also 
the parton fragmentation functions (FFs) 
describing the detected hadron in the final state. 
It is well known that the inclusion of NLA terms makes a very large effect on the 
theoretical predictions for Mueller--Navelet jet cross sections and azimuthal angle distributions. A similar behavior is expected also in our case 
of inclusive di-hadron production. This leads to a large dependence of 
predictions on the choice of the renormalization scale $\mu_R$ 
and the factorization scale $\mu_F$. Here we will take  
$\mu_R=\mu_F$ and adopt the Brodsky-Lepage-Mackenzie (BLM) 
scheme~\cite{Brodsky:1982gc} for the renormalization scale setting 
as obtained in its ``exact'' version in~\cite{Caporale:2015uva} (see Refs.~\cite{Celiberto:2016hae,Celiberto:2017ptm,Celiberto:2017ius} for more detailed studies on the use of different choices for the values of the scales).

\section{BFKL cross section and azimuthal correlations}

The process under investigation is the hadroproduction
of a pair of identified hadrons in proton-proton collisions~(Fig.~\ref{fig:di-hadron})
\begin{eqnarray}
\label{eq:process}
{\rm p}(p_1) + {\rm p}(p_2) \to {\rm h}(k_1, y_1, \phi_1) 
                              + {\rm h}(k_2, y_2, \phi_2) 
                              + {\rm X} \;,
\end{eqnarray}
where the two hadrons are characterized by high transverse momenta, 
$\vec k_1^2\sim \vec k_2^2 \gg \Lambda^2_{\rm QCD}$ 
and large separation in rapidity $Y=y_1-y_2$, 
with $p_1$ and $p_2$ taken as Sudakov vectors.
The differential cross section of the process can be presented as
\begin{equation}
\label{eq:dcs}
\frac{d\sigma}{dy_1dy_2\, d|\vec k_1| \, d|\vec k_2|d\phi_1 d\phi_2}
=
\frac{1}{(2\pi)^2}
\left[
{\cal C}_0+\sum_{n=1}^\infty  2\cos (n\phi ) \, {\cal C}_n \right] \;,
\end{equation}
where $\phi=\phi_1-\phi_2-\pi$, with $\phi_{1,2}$ the two hadrons'
azimuthal angles, while ${\cal C}_0$ gives the total
cross section and the other coefficients ${\cal C}_n$ determine 
the azimuthal angle distribution of the two hadrons. 
In order to match the kinematic cuts 
used by the CMS collaboration, we consider 
the \emph{integrated coefficients}
given by
\begin{equation}
\label{eq:Cm_int}
C_n=
\int_{y_{1,\rm min}}^{y_{1,\rm max}}dy_1
\int_{y_{2,\rm min}}^{y_{2,\rm max}}dy_2
\int_{k_{1,\rm min}}^{\infty}dk_1
\int_{k_{2,\rm min}}^{\infty}dk_2
\, \delta\left(y_1-y_2-Y\right)
\end{equation}
\[ \hspace{-6cm}
 \times  \;\; {\cal C}_n \left(y_1,y_1,k_1,k_2 \right)
\]
and their ratios $R_{nm}\equiv C_n/C_m$.
For the integrations over rapidities we consider two distinct ranges:
\begin{enumerate}
\item 
$y_{1,\rm min}=-y_{2,\rm max}=-2.4$, 
$y_{1,\rm max}=-y_{2,\rm min}=2.4$, 
and $Y \leq 4.8$, \\
typical for the identified hadron detection at LHC; \,
\item 
$y_{1,\rm min}=-y_{2,\rm max}=-4.7$, 
$y_{1,\rm max}=-y_{2,\rm min}=4.7$,
and $Y \leq 9.4$, \\
similar to those used in the CMS Mueller--Navelet jets analysis.
\end{enumerate}
As minimum transverse momenta we choose $k_{1,\rm min}=k_{2,\rm min}=5$~GeV,
which are also realistic values for the LHC. We observe that the minimum
transverse momentum in the CMS analysis~\cite{Khachatryan:2016udy} 
of Mueller--Navelet jet production is much larger, 
$k^{(\rm jet)}_{\rm min}=35$~GeV. 
The center-of-mass energy is set to $\sqrt{s} = 13$ TeV (see Ref.~\cite{Celiberto:2017ptm} for comparison with results at $\sqrt{s} = 7$ TeV).
In our calculations we use 
the PDF set MSTW 2008 NLO~\cite{Martin:2009iq} 
with two different NLO parameterizations for hadron FFs:  
AKK~\cite{Albino:2008fy} and HKNS~\cite{Hirai:2007cx}. We sum over the production of charged light hadrons: $\pi^{\pm}, K^{\pm}, p,\bar p$. 
All calculations are done in the MOM scheme. For comparison, we present results 
for the $\phi$-averaged cross section $C_0$ in the $\overline{\rm MS}$ scheme 
for $\sqrt{s} = 7, 13$ TeV and for $Y \leq 4.8, 9.4$.
In this case, we choose  natural values for $\mu_R$, {\it i.e.} 
$\mu_R = \sqrt{|\vec k_{1}||\vec k_{2}|}$, while the factorization scale is fixed to $(\mu_F)_{1,2} = |\vec k_{1,2}|$.

\section{Numerical analysis}

\begin{figure}[p]
\centering
   \includegraphics[scale=0.34,clip]{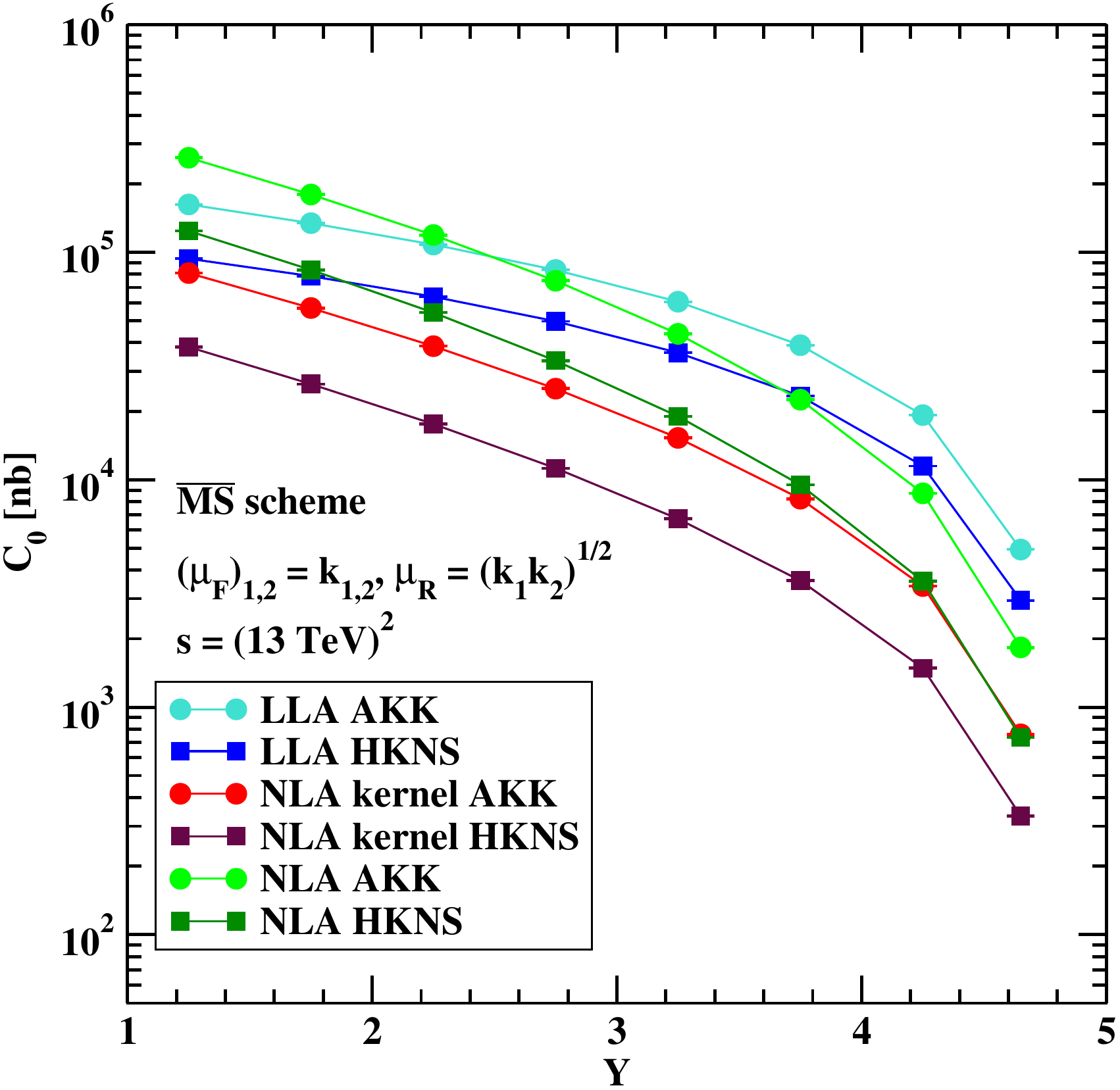}
   \includegraphics[scale=0.34,clip]{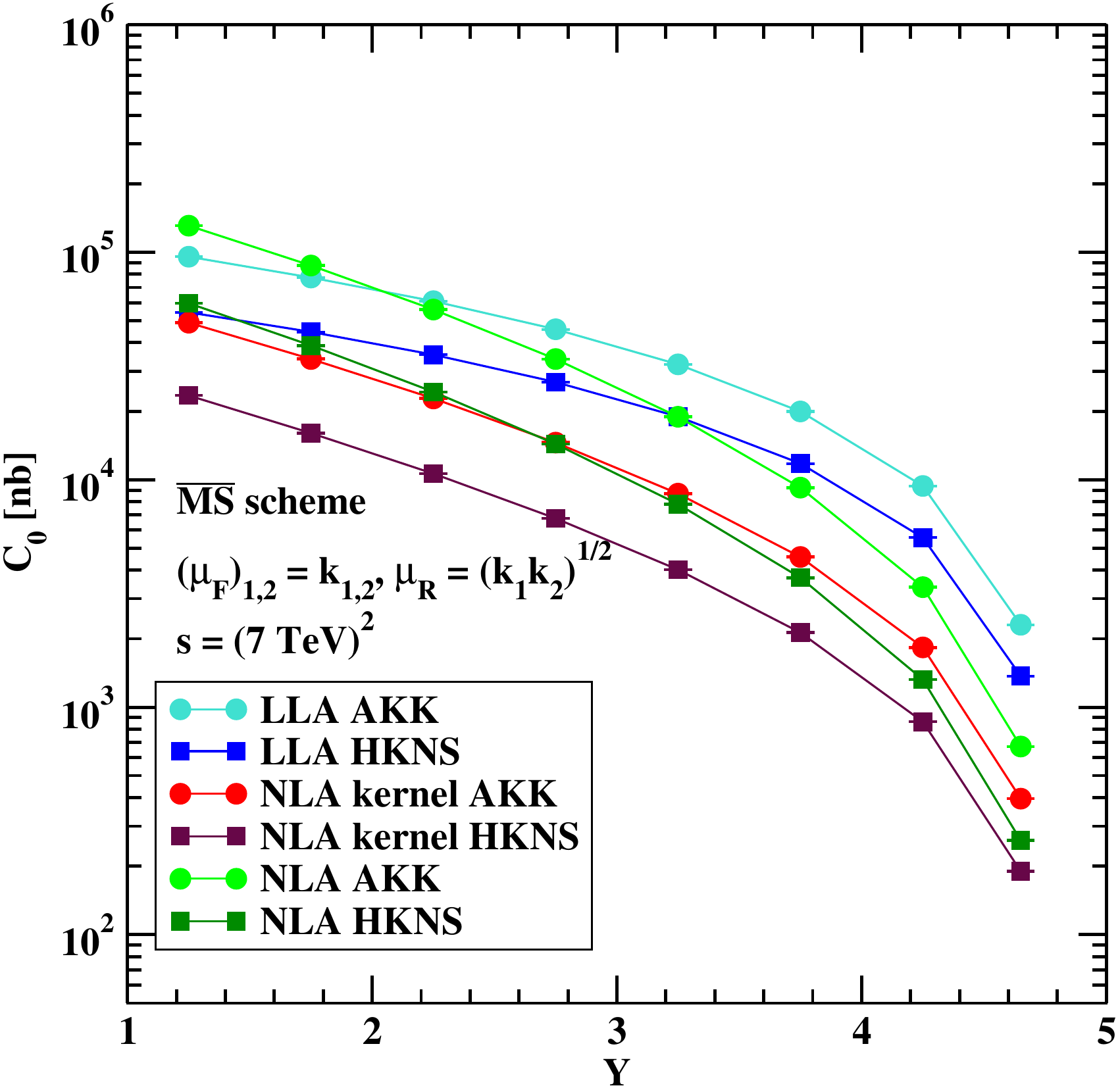}

   \includegraphics[scale=0.34,clip]{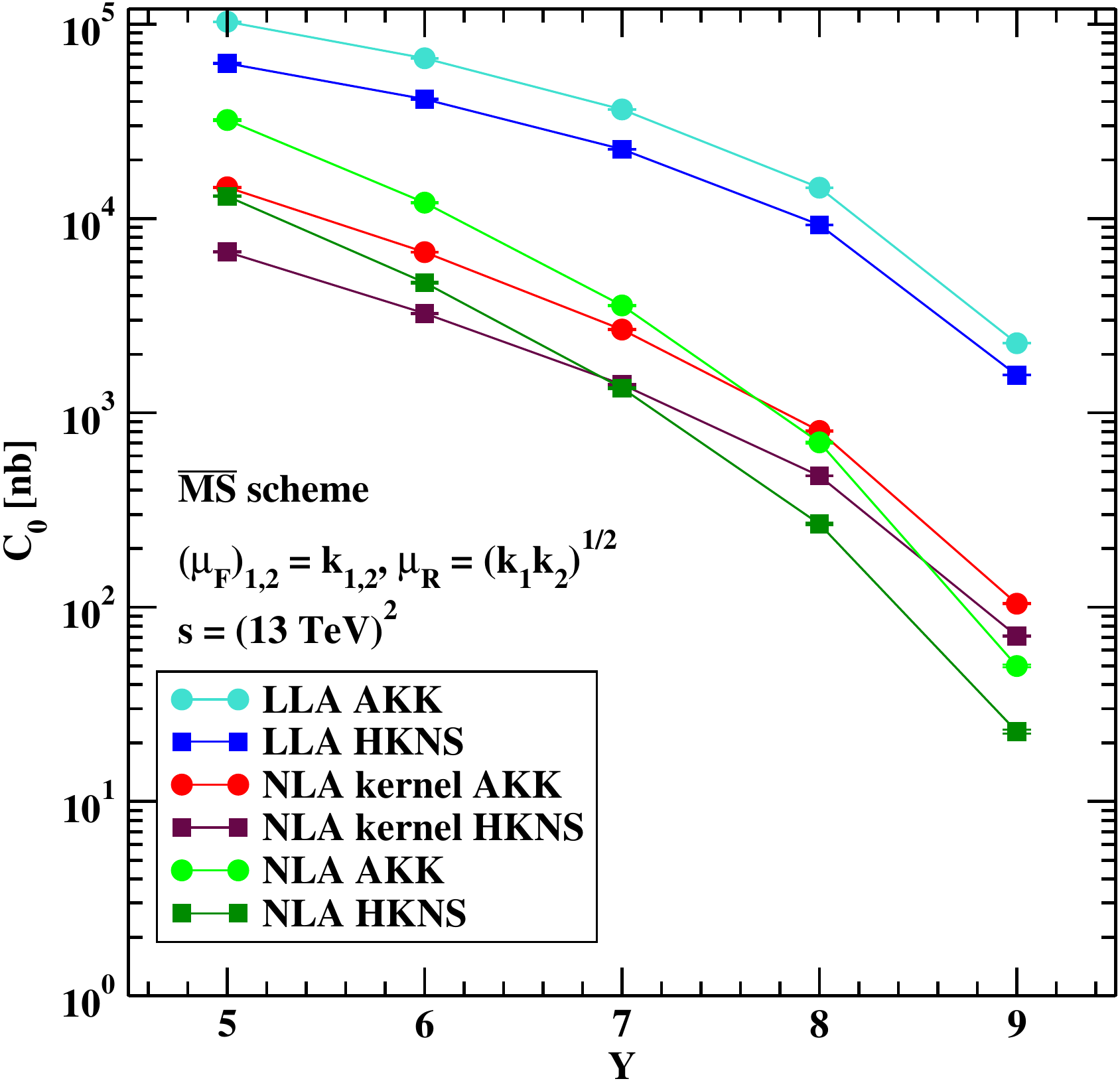}
   \includegraphics[scale=0.34,clip]{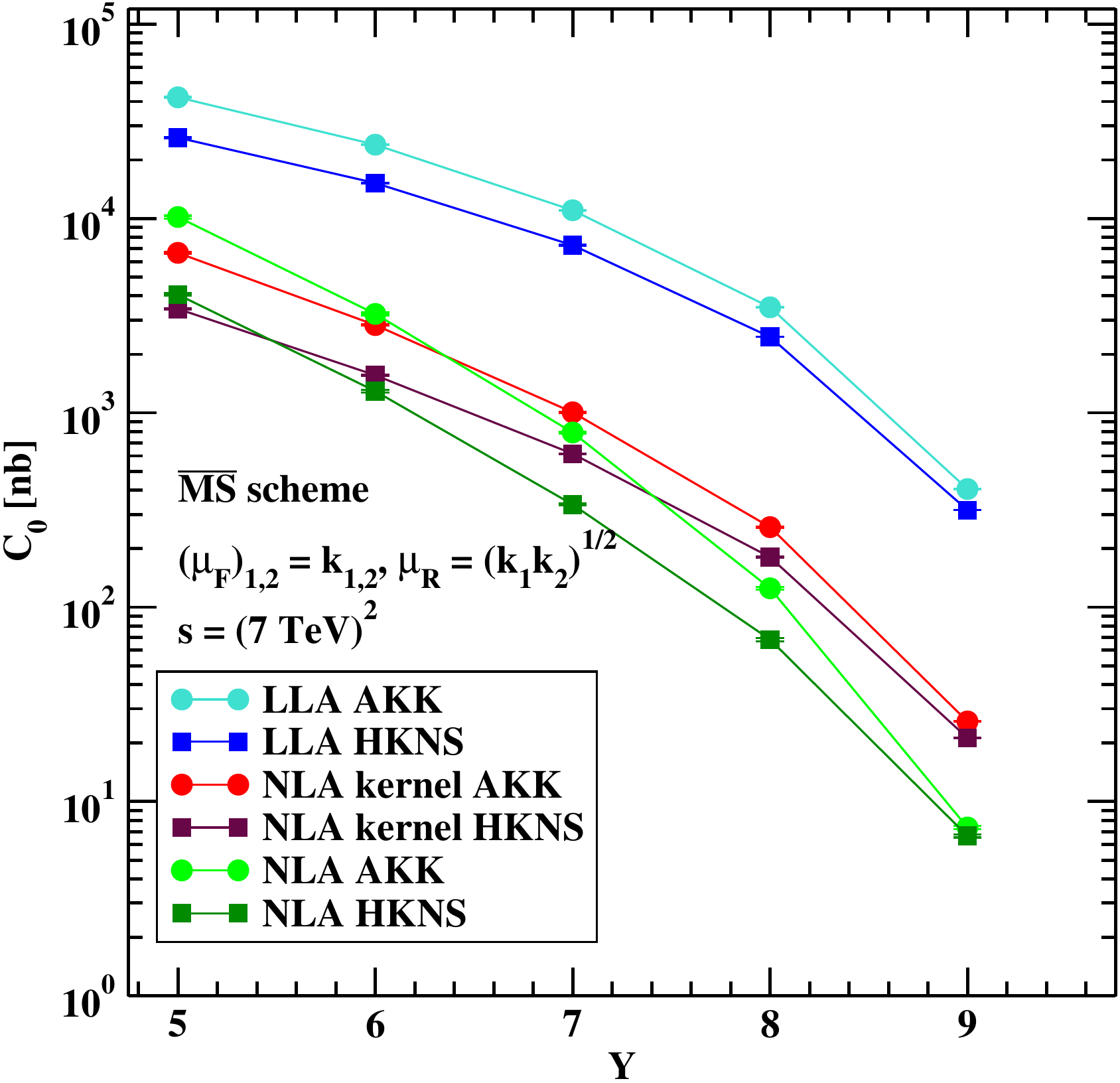}

   \caption{$Y$-dependence of $C_0$ in the $\overline{\rm MS}$ scheme at natural scales for
$\mu_R$ and $\mu_F$, $\sqrt{s} = 7, 13$ TeV, and in the two cases 
  of $Y \leq 4.8$ and $Y \leq 9.4$.  Here and in the following figure
  captions ``LLA'' means pure leading logarithmic approximation, ``NLA kernel''
  means inclusion of the NLA corrections from the kernel only, ``NLA''
  stands for full inclusion of NLA corrections, {\it i.e.} both from
  the kernel and the hadron vertices.}
\label{fig:C0MSbNS}
\end{figure}

In Fig.~\ref{fig:C0MSbNS} we present our results for $C_0$ in the
$\overline{\rm MS}$ scheme the scale settings specified above, $\sqrt{s} = 7, 13$ TeV,
and in the two cases of $Y \leq 4.8$ and $Y \leq 9.4$.
We clearly see that NLA corrections become negative 
with respect to the LLA prediction when $Y$ grows.
Besides, it is interesting to note that the full NLA approach predicts larger
values for the cross sections in comparison to the case where only NLA
corrections to the BFKL kernel are taken into account. It means that the
inclusion into the analysis of the NLA corrections to the hadron vertices
makes the predictions for the cross sections somewhat bigger and parially
compensates the large negative effect from the NLA corrections to the BFKL
kernel.

The other results we presented below are obtained using BLM in the MOM scheme, with $\mu_F$ set equal to $\mu^{\rm BLM}_R$. 

\begin{figure}[p]
\centering

   \includegraphics[scale=0.34,clip]{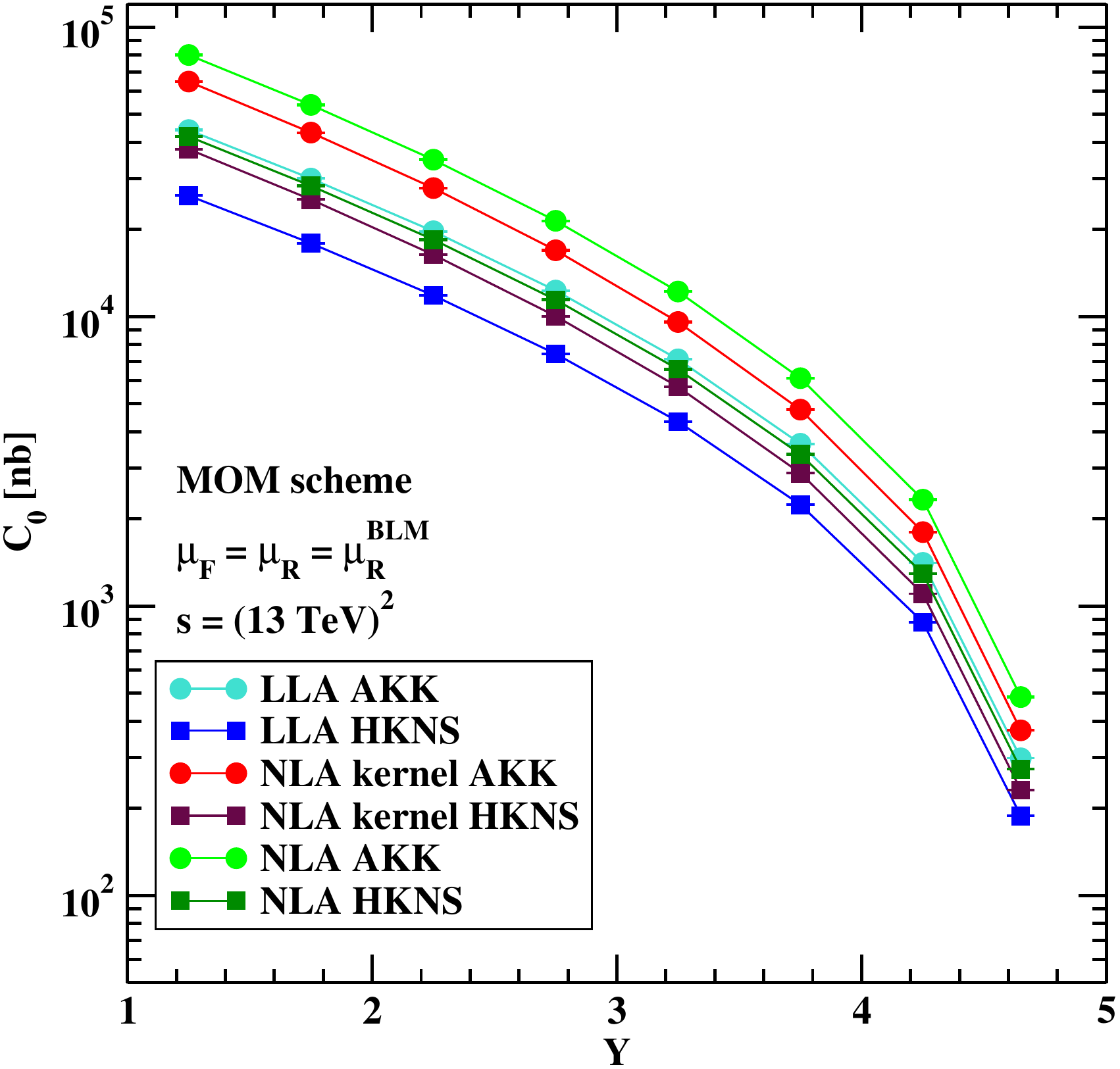}
   \includegraphics[scale=0.34,clip]{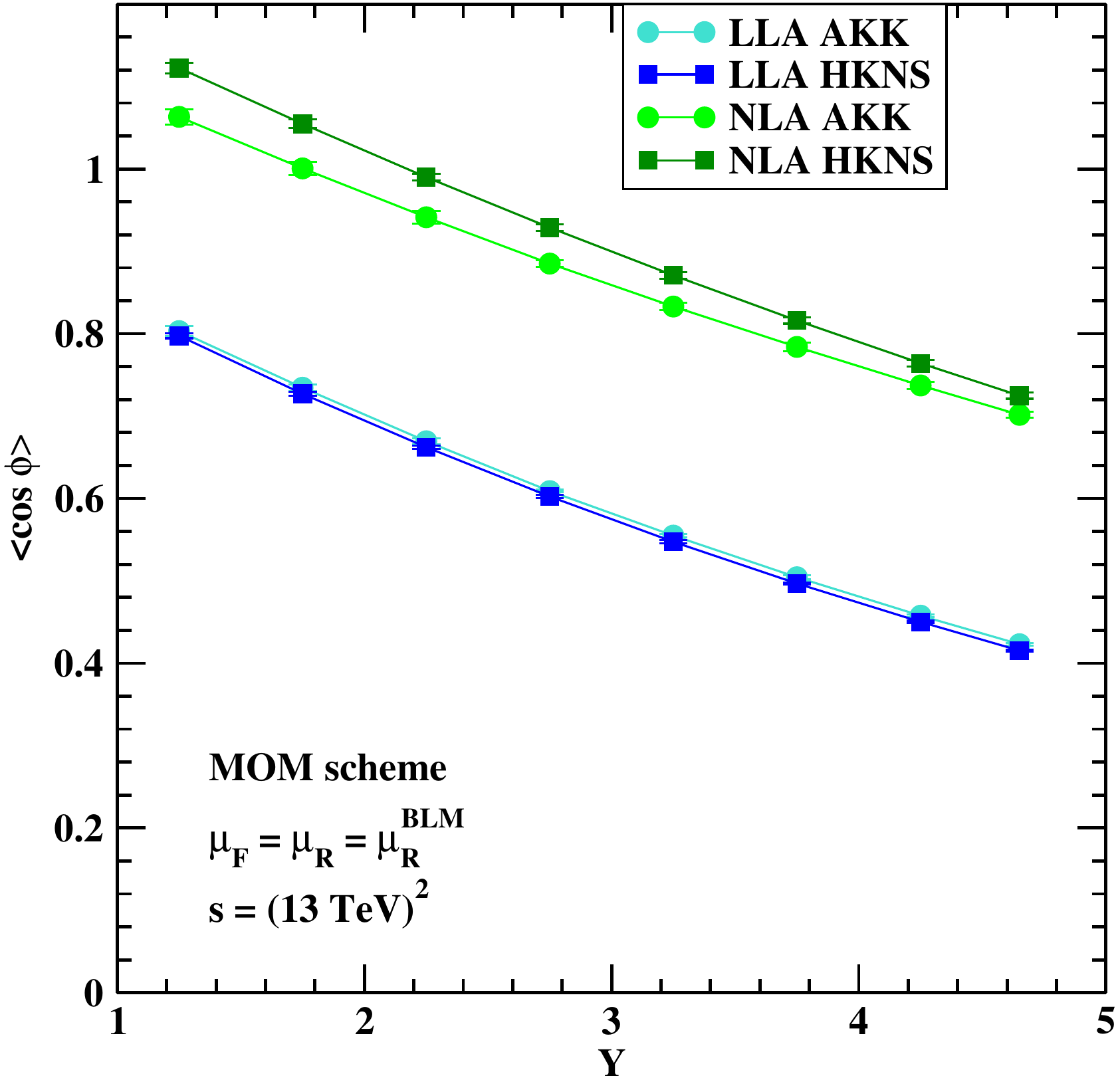}

   \includegraphics[scale=0.34,clip]{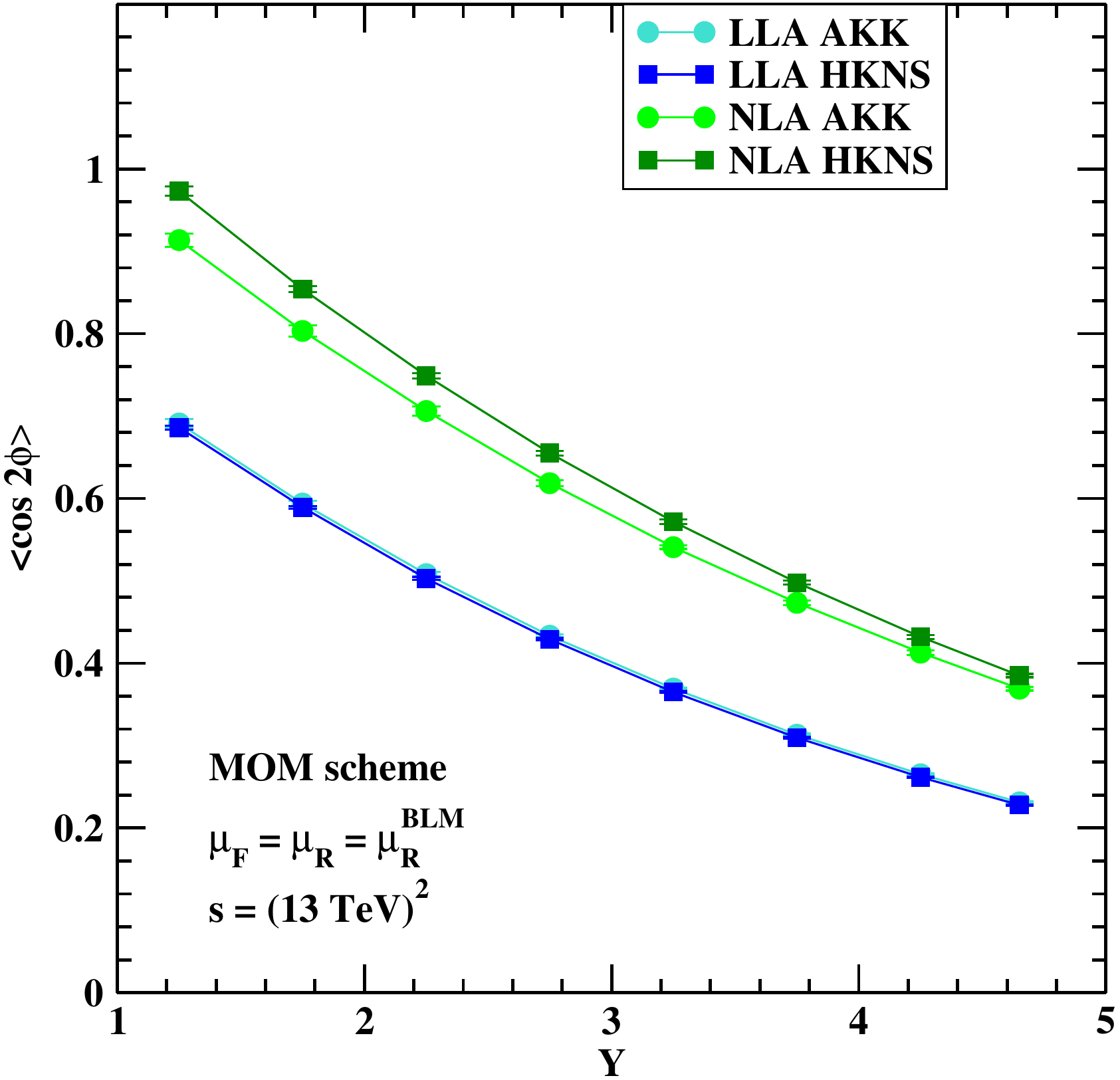}
   \includegraphics[scale=0.34,clip]{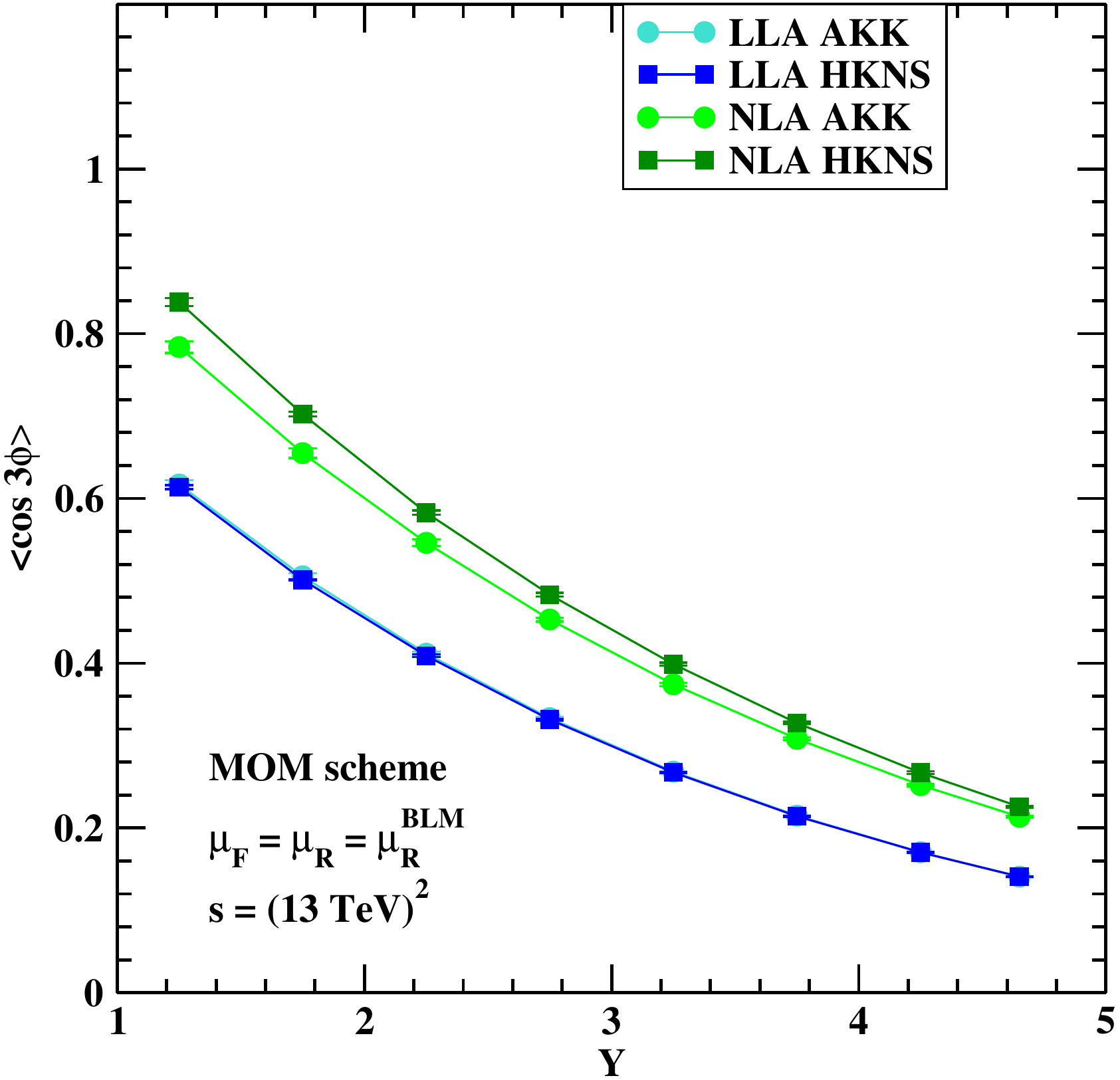}

   \includegraphics[scale=0.34,clip]{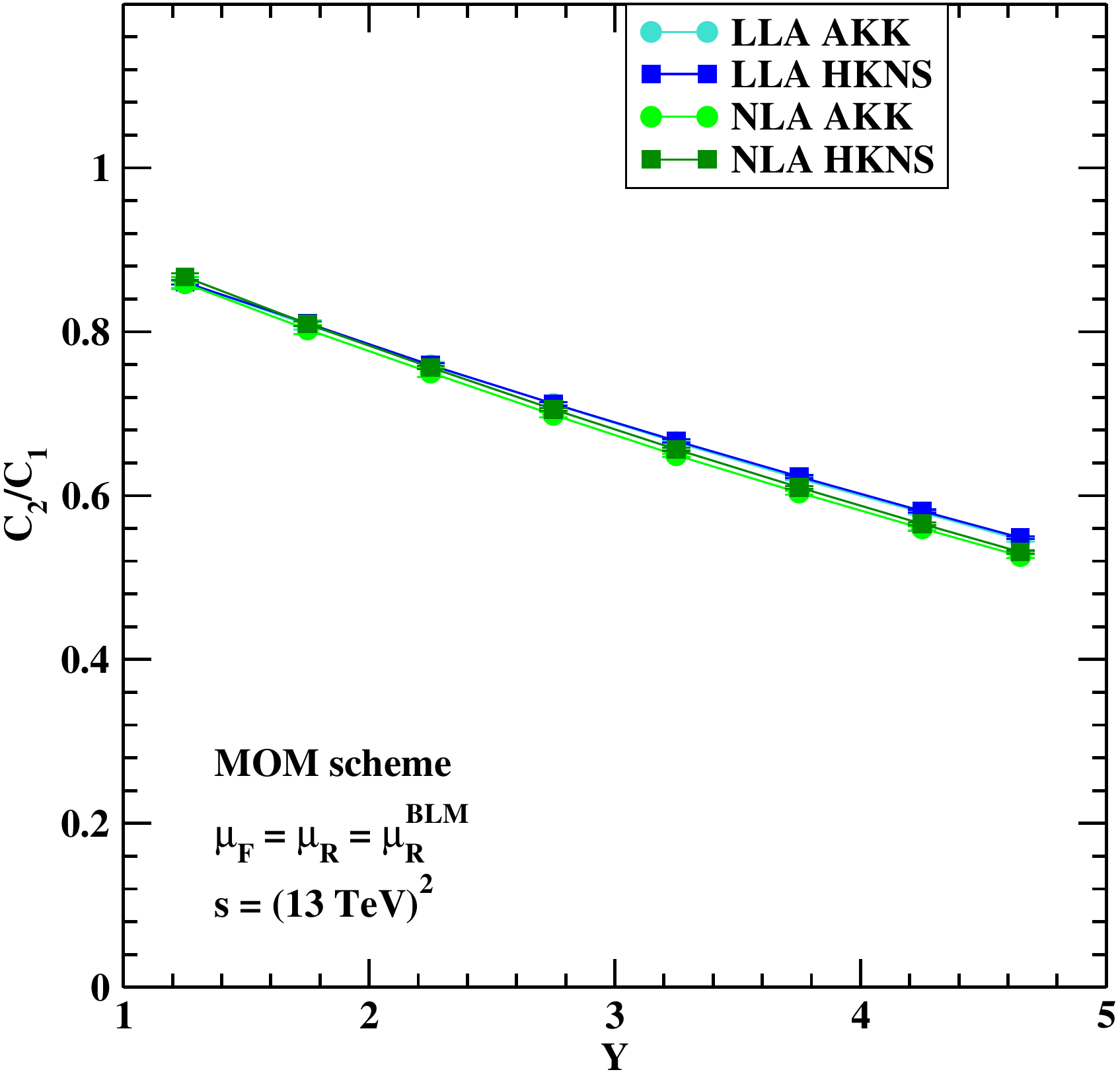}
   \includegraphics[scale=0.34,clip]{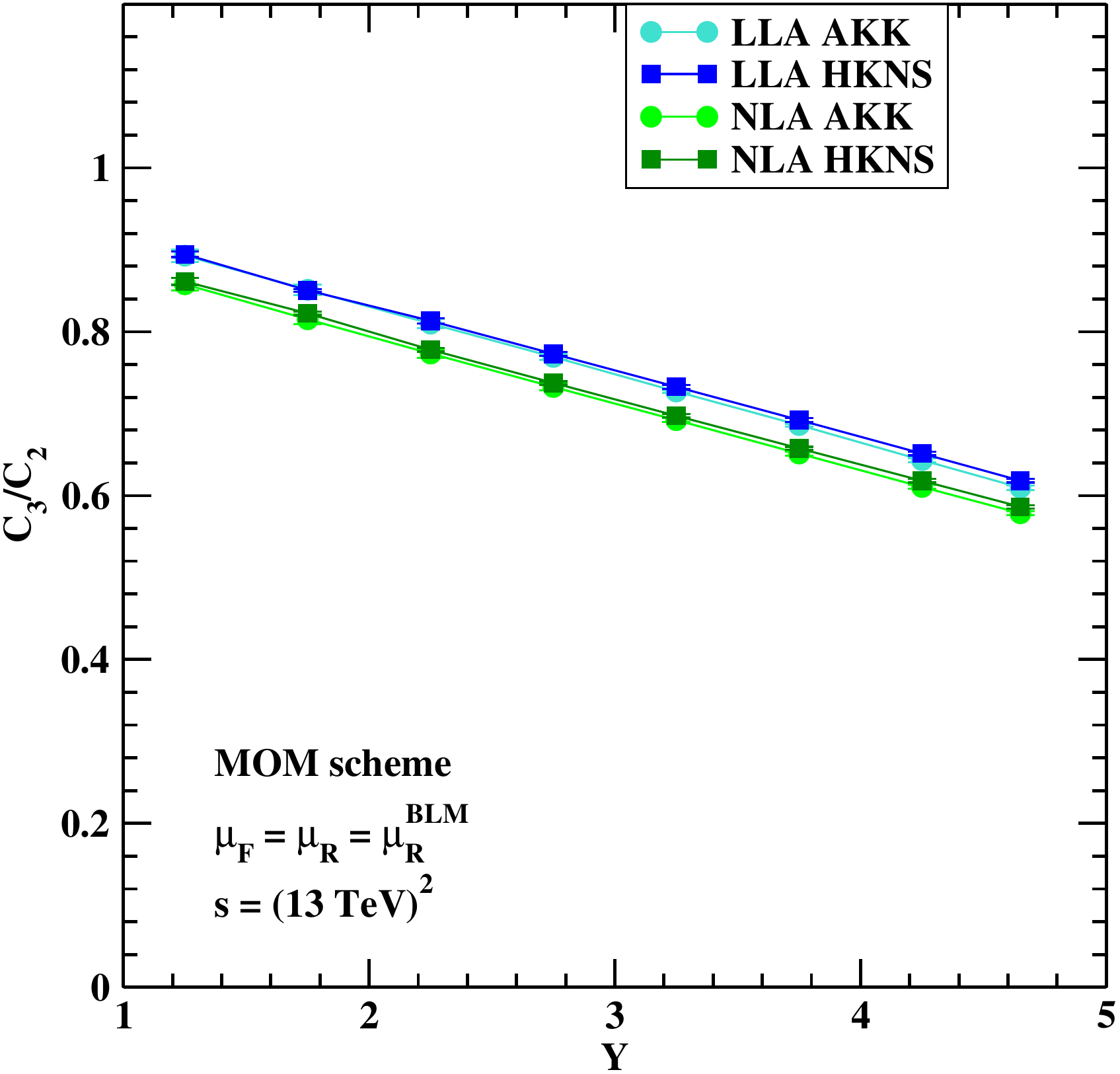}
\caption{$Y$-dependence of $C_0$ and of several ratios $C_m/C_n$ for 
$\mu_F = \mu^{\rm BLM}_R$, $\sqrt{s} = 13$ TeV, and $Y \leq 4.8$.}
\label{fig:blm13}
\end{figure}

In Fig.~\ref{fig:blm13} we present our results 
for $C_0$ and for several ratios $C_m/C_n$ at $\sqrt{s}=13$ TeV, while $Y \leq 4.8$. It is worth noting that in this case the NLA corrections 
to $C_0$ are positive, so they increase the value 
of the $\phi$-averaged cross section at all values of $Y$. 

\begin{figure}[p]
\centering

   \includegraphics[scale=0.34,clip]{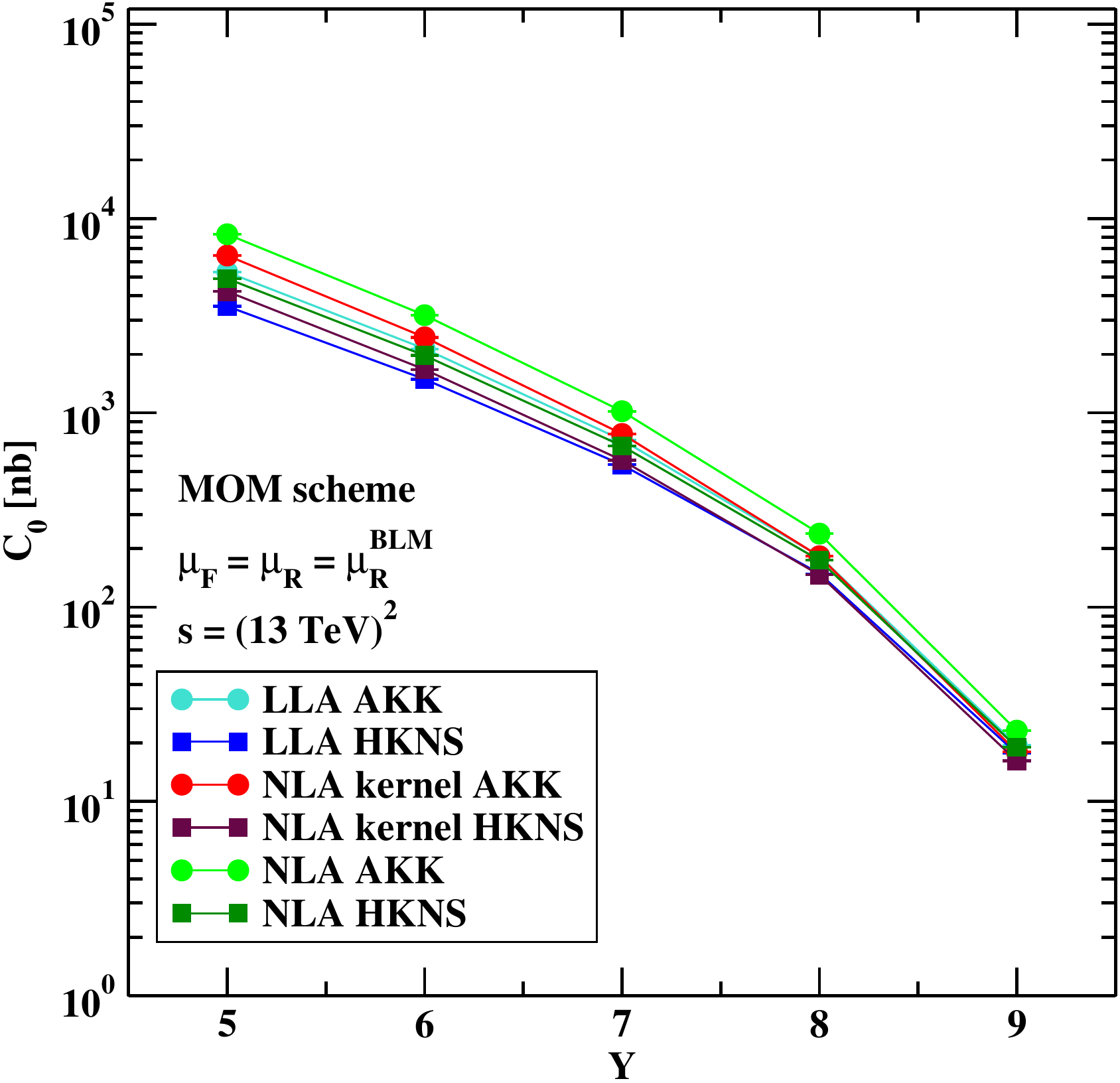}
   \includegraphics[scale=0.34,clip]{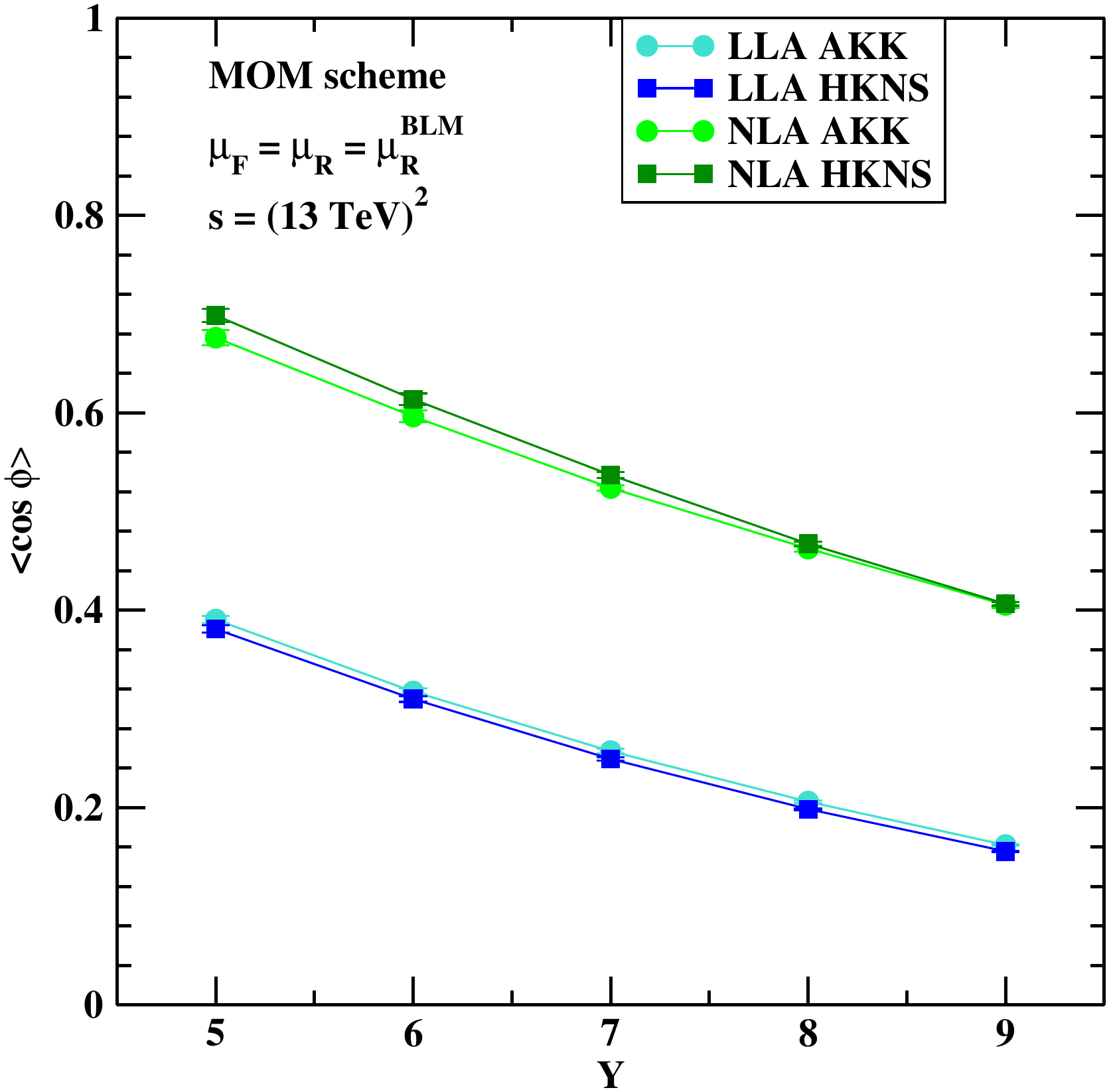}

   \includegraphics[scale=0.34,clip]{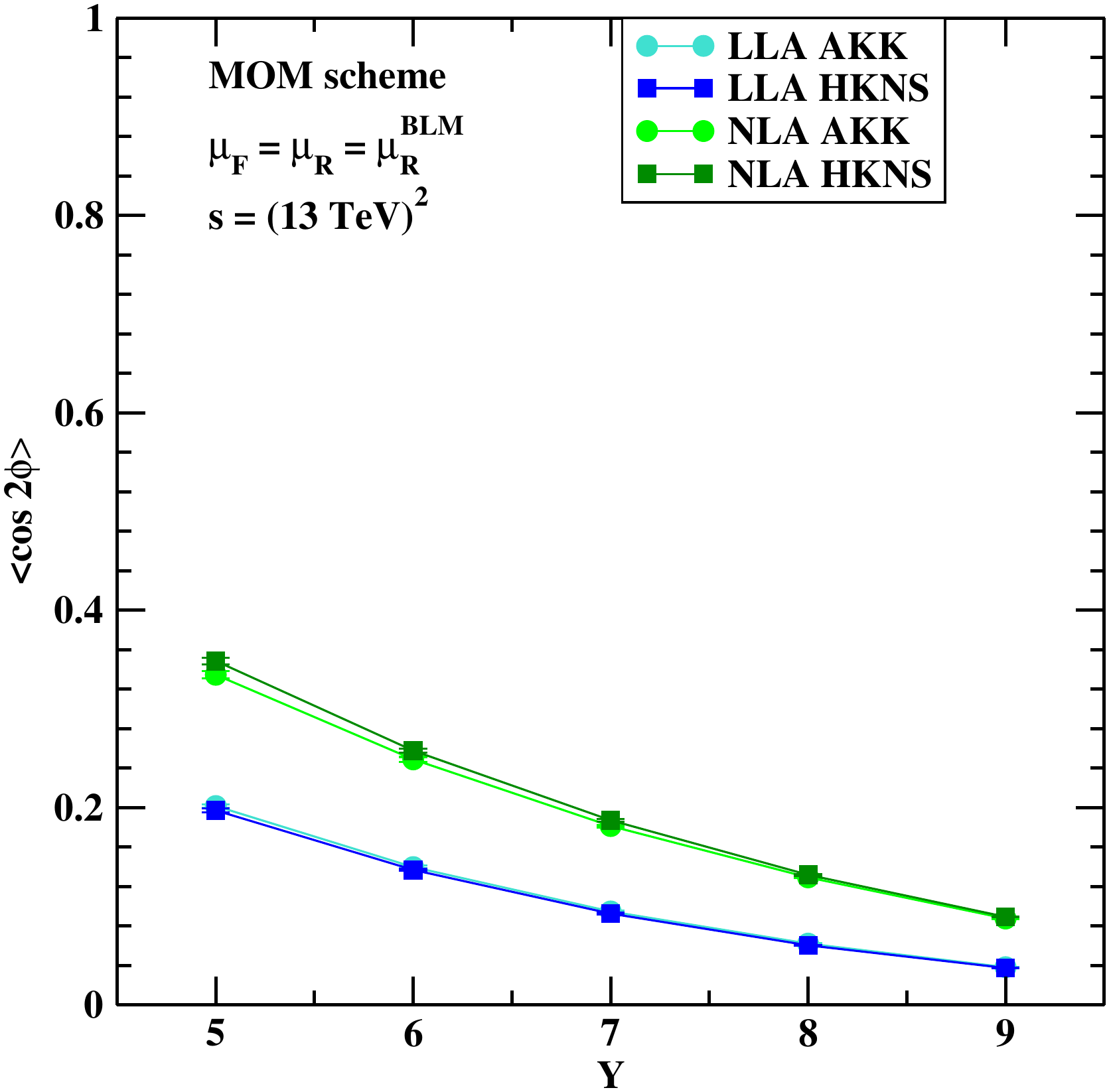}
   \includegraphics[scale=0.34,clip]{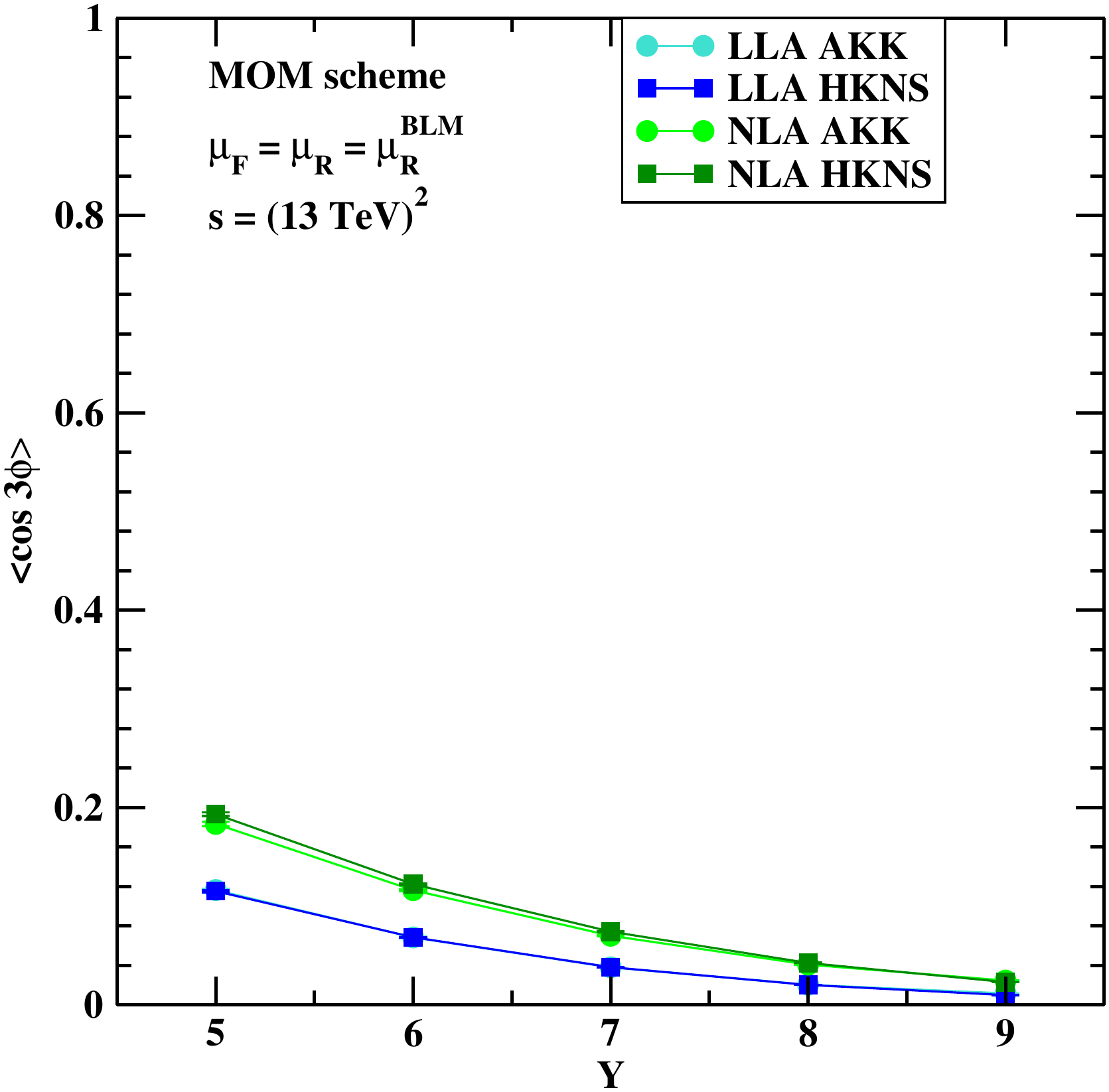}

   \includegraphics[scale=0.34,clip]{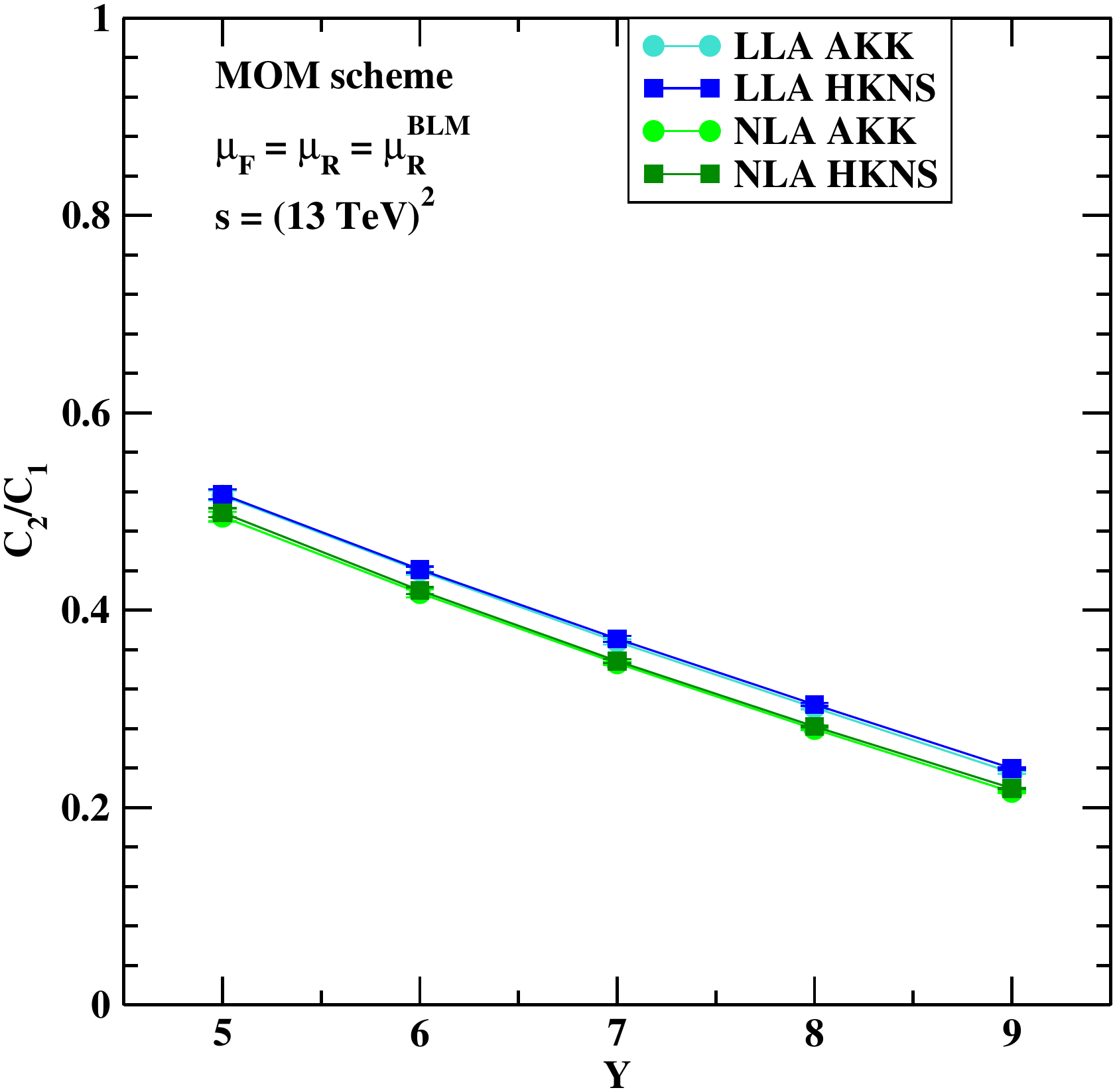}
   \includegraphics[scale=0.34,clip]{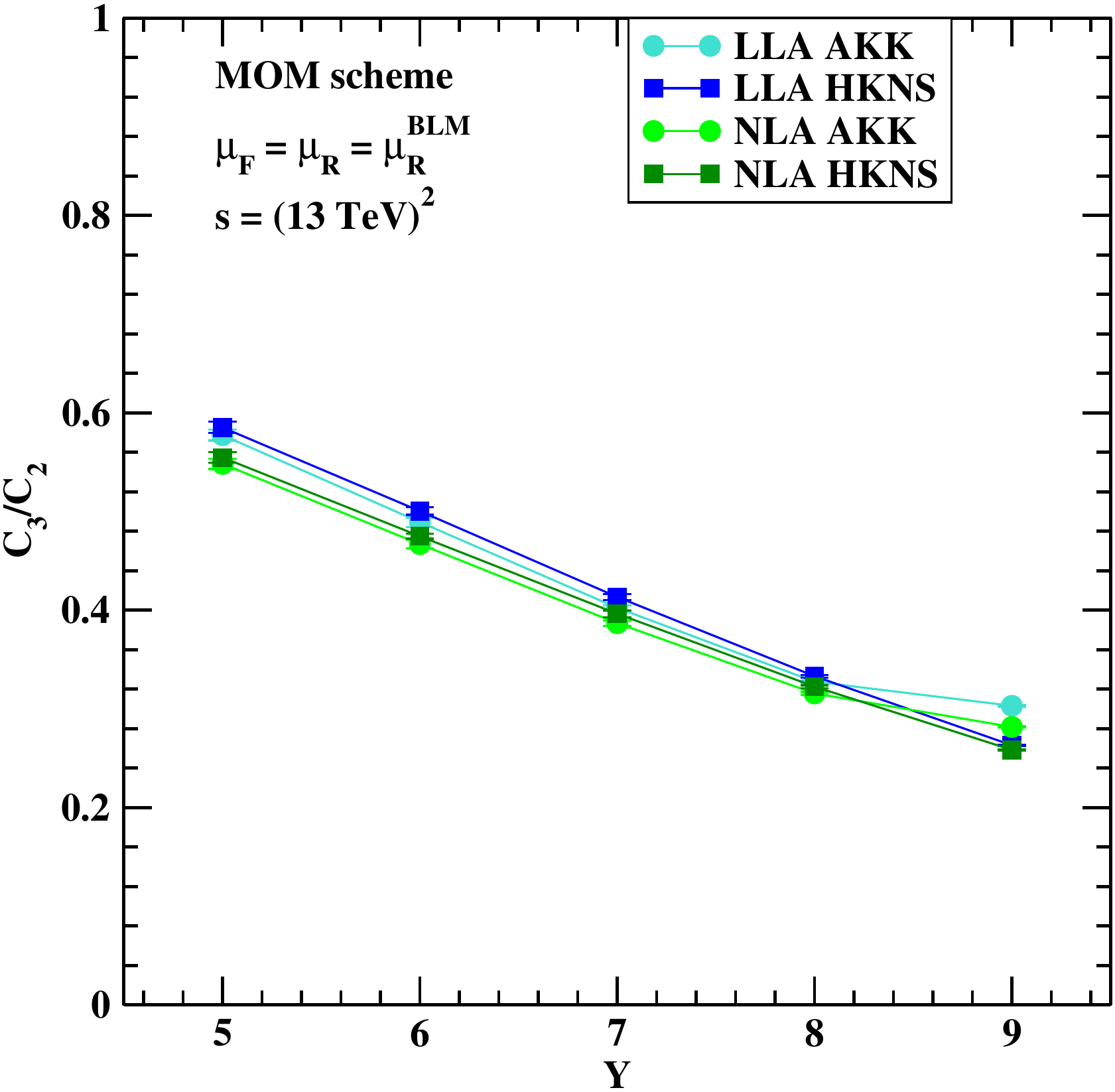}
\caption{$Y$-dependence of $C_0$ and of several ratios $C_m/C_n$ for 
$\mu_F = \mu^{\rm BLM}_R$, $\sqrt{s} = 13$ TeV, and $Y \leq 9.4$.}
\label{fig:blmLY13}
\end{figure}

In Figs.~\ref{fig:blmLY13} we present our results 
for $C_0$ and for several ratios $C_m/C_n$
at $\sqrt{s}=13$ TeV; $Y$ lies on a larger range, {\it i.e.} $Y \leq 9.4$.

\section{Summary and Outlook}

In this work we investigated the di-hadron production
process at the LHC, giving the first theoretical predictions for cross sections and azimuthal angle correlations in the full NLA BFKL approach. 
We considered the center-of-mass energy of $\sqrt s = 13$ TeV, and two different ranges for the rapidity interval between the two hadrons in the final state, $Y \leq 4.8$ and $Y \leq 9.4$, which are typical for the last CMS analyses.

The general features of our predictions for di-hadron production are
rather similar to those obtained earlier for the Mueller--Navelet jet process.
In particular, we observe that the account of NLA BFKL terms leads to much
less azimuthal angle decorrelation with increasing $Y$ in comparison to
LLA BFKL calculations.
As for the difference between the Mueller--Navelet jet and di-hadron
production processes, we would mention the fact that, contrary to the jets'
case, the full account of NLA terms leads in di-hadron production to an
increase of our predictions for the cross sections in comparison to the
LLA BFKL calculation.

We considered the effect of using different parametrization sets 
for the FFs, that could potentially give rise 
to uncertainties which, in principle, are not negligible. 

We plan to extend this study by investigating the effect 
of using asymmetrical cuts for the hadrons' transverse momenta 
as well as studying less inclusive processes where at least 
one light charged hadron is always tagged in the final state. If, together with the hadron, a forward jet is also emitted, we will
have the opportunity to study \emph{hadron-jet} correlations, which clearly enrich the
exclusiveness of the process. Another interesting reaction, which could serve as probe of the BFKL dynamics, {\it i.e.} the inclusive production of two heavy quark-antiquark pairs, separated in rapidity, in ultra-peripheral collisions at the LHC is under investigation~\cite{Celiberto:2017nyx}.

\bibliographystyle{apsrev4-1}

\newpage\newpage

\end{document}